# A Model of Cloud Based Application Environment for Software Testing

T. Vengattaraman, P. Dhavachelvan
Department of Computer Science
Pondicherry University
Puducherry, India
vengat.mailbox, dhavachelvan@gmail.com

R. Baskaran
Department of Computer Science and Engineering
Anna University
Chennai, India
baaski@cs.annauniv.edu

*Abstract*— Cloud computing is an emerging platform of service computing designed for swift and dynamic delivery of assured computing resources. Cloud computing provide Service-Level Agreements (SLAs) for guaranteed uptime availability for enabling convenient and on-demand network access to the distributed and shared computing resources. Though the cloud computing paradigm holds its potential status in the field of distributed computing, cloud platforms are not yet to the attention of majority of the researchers and practitioners. More specifically, still the researchers and practitioners community has fragmented and imperfect knowledge on cloud computing principles and techniques. In this context, one of the primary motivations of the work presented in this paper is to reveal the versatile merits of cloud computing paradigm and hence the objective of this work is defined to bring out the remarkable significances of cloud computing paradigm through an application environment. In this work, a cloud computing model for software testing is developed.

## I. INTRODUCTION

Cloud computing is a recent evolution of distributed computing paradigm which can support on-demand service sharing with higher level of flexibility and dynamic scalability. Flexibility and scalability in cloud computing environment can be accomplished through load balancing of application instances running separately on a variety of operating systems and connected through Web services [1]. Cloud computing provide Service-Level Agreements (SLAs) for guaranteed uptime availability. Furthermore, cloud computing make use of a model for enabling convenient and on-demand network access to the distributed and sharable computing resources which are configurable [2] [4]. The greatest advantage of the cloud computing is the user need not have knowledge of the technology and control over the infrastructure in the cloud environment.

Though the cloud paradigm holds its potential status in the field of distributed computing, cloud platforms aren't yet at the center of most people's attention. More specifically, still the researchers and practitioners community has fragmented and imperfect knowledge on the principles and techniques of cloud computing paradigm. Due to these, other paradigms are seems to be better, but this scenario will be changed shortly on behalf of the real attractions of cloud based computing, including scalability and cost effective services [2-3]. So it is the responsibility of the researchers to make the practitioners to realize the flavor of cloud computing environment. In this context, one of the primary motivations of the work presented in this paper is to reveal the versatile merits of cloud computing paradigm and hence the objective of this work is defined to bring out the remarkable significances of cloud computing paradigm through an application environment. This development is aimed to a broad extent and its significances are realized in a limited scope, which can be enhanced to its appropriate extent.

Once a decision has made to implement a cloud environment, the designer is supposed to adopt a standard procedure for implementing the same. But at present there is no such standard procedure or guidelines are available for building the cloud computing environment and hence here the services are implemented using web service concepts. The technology of web service based systems has generated lots of interest among the user and developer in the recent years through its ability as a new paradigm for conceptualizing, designing, and implementing software systems. This characteristic is particularly attractive for creating software that operates in environments that are distributed and open, such as the internet and this property increases the need for web service systems that consist of services that communicate in a peer-to-peer fashion is becoming evident. Central to the design and effective operation of such web services are a core set of issues and research questions that have been studied over the years by the distributed community [10-12]. They are large-scale systems and the research community aims at web service collaboration to achieve their functions in a highly flexible manner. But unfortunately, in the existing systems, too few are described with sufficient details to adopt them for real world problems [8-9]. The solution is to construct and use objective specific frameworks that are best suited for the given problem space. From this perspective here, an objective specific web service based cloud framework for software testing is developed.

From the cloud management point of view, a critical challenge in creating effective clouds is making them robust against potential failures. In few cases, system failures may causes the poor performance of the application though it has been well developed and deployed. Thus, in addition to the functionality features, the application should able to coordinate with the system that is handling the particular application. On other hand, the cloud should also be implemented with such capabilities to coordinate with the applications. So, appropriate failure or exception handlings in the cloud framework are not only to improve the quality of the system in terms of availability, also to improve the performance of the application domain. In this view, the possible set of exceptions and exception handling mechanism are also described in order to assist in management of clouds presented in this paper. The mechanisms can also be considered as generic and may be applied wherever they required.

The organization of paper is as follows: Section 2 defines the proposed enhanced version of web service for software testing as an application domain. Section 3 explains the experimentation methodology to validate the proposed system in two perspectives; performance assessment of the framework w.r.t. the application environment at the service level and competence assessment of the framework w.r.t. the computing environment management in terms of cloud computing performance attributes. Section 4 presents the quantitative statistical analysis over the results obtained from the experiments and the Section 5 offer the concluding remarks and directions of future enhancements.

## II. PROPOSED SYSTEM

### A. Introduction of Cloud

Cloud computing is a recent evolution of distributed computing paradigm that uses the internet and central remote servers to maintain data and applications and thereby allows the users to access technology-enabled services from the Internet without knowledge of, expertise with, or control over the technology infrastructure that supports them. Dynamism and resource sharing are the key architectural indicators of cloud computing and this technology allows much more efficient computing by centralizing storage, memory, processing and bandwidth. In general, the cloud services can be grouped into three broad categories [1] [3-7]:

*Software as a service (SaaS):* It is a kind of application runs exclusively in the cloud. The services are typically hosted and managed in their own data center and make it available over the Web.

*Platform as a Service (Paas):* These categories of services are the on-premises applications that are to be built on shared platform in the cloud. This system facilitates development and deployment of applications without the cost and complexity of buying and managing the underlying infrastructure and enhances the on-premises application functionalities by accessing the application specific services provided in the cloud. They can be viewed as the set of attached services useable only by a particular application.

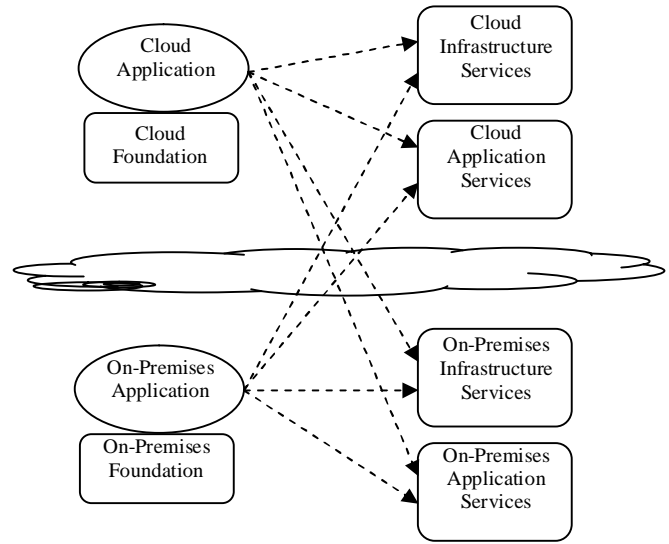

Figure 1. Interconnection between On-Premises Applications and Cloud Applications

*Infrastructure as a Service (PaaS):* Infrastructure as a Service (IaaS) is the delivery of hardware and associated software as a service. A cloud platform provides cloud-based services for creating applications rather than building their own custom foundation, they can be built on a cloud platform.

The general interaction between the on-premises applications and cloud services [1] are described in the Fig 1. The work presented in the paper is based on the PaaS category, which includes the application services of on-premises and cloud environments.

### B. Proposed Model Development

In modeling the proposed System, it will be easier to understand the model by expressing it in the product based definition form. In this form, the model is used to define different sets of Testing Services with respect to the product, which is to be tested. In all the cases, the second category is the subset of the former one. From the product perspective, the proposed system can be defined in terms of product specific services.

Definition-1: Let '$A_i$' be the set of Clouds needed for testing the product '$P_i$' and with respect to $P_i$, 'A' can be defined as in the Equation (1)

$$A_{P_i} = \begin{cases} (sm_i, tc_{i1}, tc_{i2}, tc_{i3}, tc_{i4}, \ldots tc_{iy}), \\ tc_{i1} = (ts_{i(11)}, ts_{i(12)}, \ldots ts_{i(1K_{i1})}), \\ tc_{i2} = (ts_{i(21)}, ts_{i(22)}, \ldots ts_{i(2K_{i2})}), \\ . \\ . \\ tc_{iy} = (ts_{i(y1)}, ts_{i(y2)}, \ldots ts_{i(yK_{iy})}) \end{cases} \quad (1)$$

where,

- '$sm_i$' is the Service Manager of the product '$P_i$' belongs to the On-Premises Application Services and 'H' is the number of products to be tested simultaneously and then $0 < i \leq H$.
- '$tc_{ij}$' is the one of the Testing Cloud of '$A_i$' and '$tc_{i(ju)}$' is the one of the Testing Services of '$tc_{ij}$' and then $0 < j \leq y$, $0 < u \leq k_{ij} - 1$.
- 'y' is the number of Clouds and also the number of different Testing Clouds required for the product '$P_i$'.
- '$K_{ij}$' refers to the maximum number of Testing Services of '$a_{ij}$' and '$K_{ij}$' refers to the total number of Testing Services available in the particular Testing Clouds 'j' of the product '$P_i$'.

Since this cloud framework provides, scalar type testing environment, at any instant, $A_{i_1} \cap A_{i_2} = \phi$, where, $0 < i_1, i_2 \leq H$ and $i_1 \neq i_2$. i.e. at any specific service duration, neither a single Testing Cloud nor a Testing Services can be shared by more than one product simultaneously. The logical layout of the proposed cloud framework for software testing is shown in the Fig 2.

Generally there are three sets of components as Service Manager $(\{sm_1, sm_2, \ldots sm_z\})$, Testing Clouds $(\{tc_1, tc_2, \ldots tc_x\})$ and the clones of Testing Services $(\{ts_{11}, ts_{12}, \ldots ts_{1K_1}\} \ldots \{ts_{x1}, ts_{x2}, \ldots ts_{xK_x}\})$ each of which runs locally on different machines in a network. The logical links between the services refers to the service level dependency among the services in the framework. Such a dependency can represent the fact that one service depends on another for a goal to be fulfilled, task to be performed, or a resource to be made available.

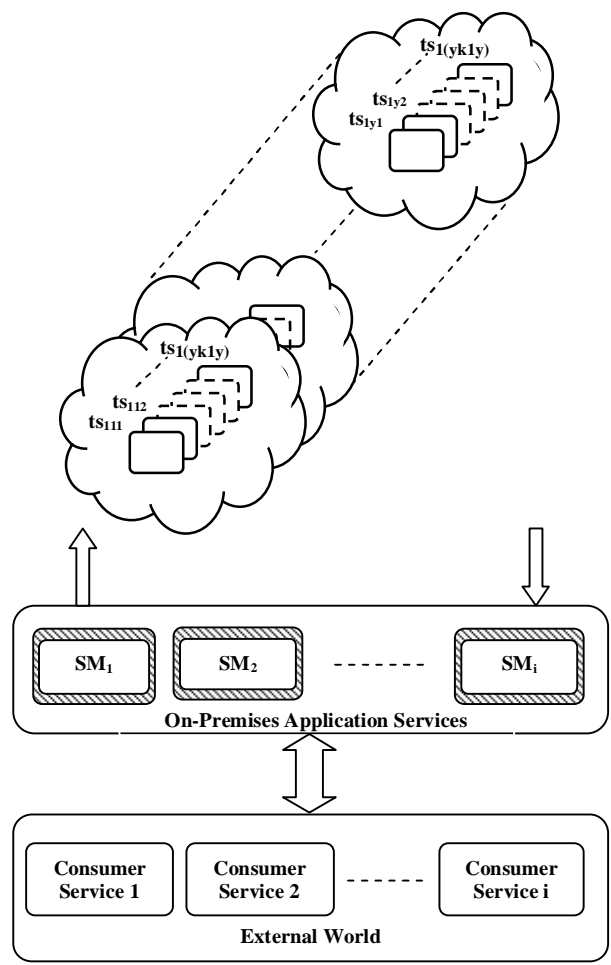

Figure 2. Proposed Cloud Environment

In specifying the proposed system, it has been found that it is highly desirable to explain the model in a layered approach; there are two types (levels) of services: On-Premises Application Services and Cloud Application Services. The cloud application services layer is the dynamic one, which will be generated at the time of the testing processes. The overall support is composed of units, which specialize in supporting different testing services.

The Service Manager is responsible for supervision and coordination of the core activities of respective layer entities of the overall environment. After getting the assignment from the user, the Service Manager ($SM_i$) will form the Testing Clouds and can define the set $\{tc_1, tc_2, \ldots tc_x\}$. The process of defining the set of required clouds depends on the testing techniques required for any product '$P_i$'. Then, based on the load, each Testing Cloud will select one or more additional Testing Services of its respective techniques and define the sets of Testing Services. These additional services are to be referred to as secondary services of Service Manager.

Each cloud is responsible for providing a specific output standard to which the task output must conform. The output standard depends on the task type and the required output properties such as number of test cases, defects found, time spent for testing, etc. The input to the Consumer Service 'CS$_i$' is from the tester and it includes the set of the testing product, time specification for testing, defect detector estimations and the specification about the required testing techniques. This is transferred to the Service Manager 'SM$_i$' of 'CS$_i$'. The Service Reception receives the input from the CS$_i$ and analyze the specification about the required testing techniques. Based on the testing service specifications, the appropriate set of Clouds can be defined and identified through the negotiation process in the Cloud Application Services. Upon the agreement of the negotiation process 'K$_y$' Clouds are formed.

Then the assignments will be distributed to the identified set of Testing Cloud and their outputs can be obtained for integration. The Environmental Test Reports from the identified testing Clouds will be integrated in the testing management block and then passed to the external world (CS$_i$) through the service reception of SM$_i$. The Service Manager is responsible for all types of co-ordination activities in this system. The initial input to the Testing Cloud 'tc$_{ij}$' is from the Testing Application Management. It includes the set of the testing techniques, time specification for testing, defect detector estimations and the estimated complexity of the product. This is transferred to the Testing Clouds 'tc$_{ij}$' with a set of estimated values on average size of the modules of the product and the predicted values of total number of test cases to be built by tc$_{ij}$, average size of the test case and average time required for generating and executing an unit test case.

Service Manager will define the mode of load distribution and it is based on the input from the Consumer Service. This is also responsible for defining the number of Testing Services that are needed to generate. Moreover processes of registration, load distribution and collection of results from the Testing Services are to be done. At the same time it will distribute the load to the Testing Clouds of same tc$_{ij}$. The results from the Testing Clouds i.e. Environmental Partial Test Reports (EPTRs) from the Testing Services will be processed in the Testing Application Management. Then the generated Environmental Test Report (ETR) of tc$_{ij}$ will be transferred to the Consumer Service CS$_i$.

### III. CONCLUSION

In this work, a model of cloud computing environment for software testing has been developed with different types of clouds. This work is one of the very few examples of cloud frameworks constructed for problem solving in software engineering field, which is being explicitly integrated as a service in a distributed network environment and validated with multiple folds; framework assessment with respect to the application environment and framework assessment with respect to the computing environment management. From the software engineering perspective, it demonstrated how a new web service paradigm enables legacy software to be incorporated in a relatively straightforward manner under any kind of circumstances.